\documentclass[
reprint,
superscriptaddress,
 amsmath,
 amssymb,
 aps,
prl,
]{revtex4-1}

\usepackage{graphicx}
\usepackage{dcolumn}
\usepackage{bm}
\usepackage{physics}
\usepackage{siunitx}

\usepackage{booktabs}
\usepackage{multirow}
\usepackage{makecell} 
\usepackage{mathtools} 
\DeclareSIUnit\angstrom{\text{Å}}
\usepackage{braket}
\usepackage{xcolor}
\usepackage[colorlinks=true, linkcolor=blue, citecolor=blue, urlcolor=blue]{hyperref}
\usepackage{array}

\begin{document}
\setcounter{secnumdepth}{1}

\title{Zeeman Spectroscopy of Vacancy-Charge-Compensated \texorpdfstring{Er\(^{3+}\)}{Er3+} Sites in CaWO\texorpdfstring{\(_4\)}{4} under Vector Magnetic Fields}

\author{Fabian Becker}
    \affiliation{TUM School of Computation, Information and Technology, Technical University of Munich, 80333 Munich, Germany}%
    \affiliation{Walter Schottky Institute, Technical University of Munich, 85748 Garching, Germany}%
    \affiliation{Munich Center for Quantum Science and Technology (MCQST), 80799 Munich, Germany}%
\author{Sudip KC}
    \affiliation{TUM School of Computation, Information and Technology, Technical University of Munich, 80333 Munich, Germany}%
    \affiliation{Walter Schottky Institute, Technical University of Munich, 85748 Garching, Germany}%
    \affiliation{Munich Center for Quantum Science and Technology (MCQST), 80799 Munich, Germany}%
\author{Lorenz J. J. Sauerzopf}
    \affiliation{TUM School of Computation, Information and Technology, Technical University of Munich, 80333 Munich, Germany}%
    \affiliation{Walter Schottky Institute, Technical University of Munich, 85748 Garching, Germany}%
    \affiliation{Munich Center for Quantum Science and Technology (MCQST), 80799 Munich, Germany}%
\author{Tim Schneider}
    \affiliation{TUM School of Computation, Information and Technology, Technical University of Munich, 80333 Munich, Germany}%
    \affiliation{Walter Schottky Institute, Technical University of Munich, 85748 Garching, Germany}%
    \affiliation{Munich Center for Quantum Science and Technology (MCQST), 80799 Munich, Germany}%
\author{Luis Risinger}
    \affiliation{TUM School of Computation, Information and Technology, Technical University of Munich, 80333 Munich, Germany}%
    \affiliation{Walter Schottky Institute, Technical University of Munich, 85748 Garching, Germany}%
    \affiliation{Munich Center for Quantum Science and Technology (MCQST), 80799 Munich, Germany}%
\author{Christian Schmid}
    \affiliation{TUM School of Computation, Information and Technology, Technical University of Munich, 80333 Munich, Germany}%
    \affiliation{Walter Schottky Institute, Technical University of Munich, 85748 Garching, Germany}%
    \affiliation{Munich Center for Quantum Science and Technology (MCQST), 80799 Munich, Germany}%
\author{Kai M\"uller}
    \affiliation{TUM School of Computation, Information and Technology, Technical University of Munich, 80333 Munich, Germany}%
    \affiliation{Walter Schottky Institute, Technical University of Munich, 85748 Garching, Germany}%
    \affiliation{Munich Center for Quantum Science and Technology (MCQST), 80799 Munich, Germany}%

\date{\today}

\begin{abstract}
We present polarization-resolved optical absorption measurements on Er$^{3+}$ ions in CaWO$_4$ under vector magnetic fields, focusing on charge-compensated sites arising from local Ca$^{2+}$ vacancies. While the known axial Er$^{3+}$ site displays a single symmetric Zeeman-split transition pattern consistent with S$_4$ symmetry, two additional sites exhibit more complex spectral behavior, including sets of transitions that interchange under $\SI{90}{\degree}$ crystal rotations—evidence of reduced, rhombic-like symmetry. From these polarization- and temperature-dependent spectra, we extract effective g-factors. Our findings are corroborated by electron paramagnetic resonance measurements and support a model of multiple inequivalent Ca$^{2+}$ vacancies around Er$^{3+}$ sites in the host lattice. This detailed characterization contributes to understanding defect-engineered rare-earth sites for quantum information applications.
\end{abstract}
\maketitle
\section{Introduction} \label{S_Introduction}
In the past years, the interest in CaWO$_4$ research has been and still is experiencing a renaissance driven by its function as a quantum host \cite{LeDantec.2021, Ourari.192023, Rancic.2022, Uysal.6102024, Wang.2023, Billaud.2025}. Especially its long spin coherence time with naturally abundant host materials is outstanding \cite{LeDantec.2021}. Furthermore, the incorporation of Er$^{3+}$ gains special attention due to its emission in the telecom low-loss window, which is essential for any fiber-based quantum network \cite{Reiserer.2022}. For the Er:CaWO$_4$ system, indistinguishability \cite{Ourari.192023} and spin-photon entanglement \cite{Uysal.6102024} measurements showed first application proofs of concepts.\\
For the incorporation of rare-earth ions (REIs) into CaWO$_4$, it is known that charge compensation takes place to incorporate a triply ionized REI at the double ionized Ca site \cite{Billaud.2025, Mims.1965, Nassau.1963, Nassau.1963b}. In this process, the charge compensation is mainly of a long-range nature, such that the main or axial symmetric incorporation occurs. However, short-range compensation sites without axially symmetric incorporation are also observed with the long-range/short-range ratio decreasing with increasing doping amount of triply ionized species \cite{Nassau.1963}. For Er:CaWO$_4$, we identified in a previous optical study at least three additional environments which we attributed to these additional short-range compensation sites \cite{Becker.2025}. To our knowledge, a side note in an Electron Paramagnetic Resonance (EPR) study from 1968 reported these additional sites for Er:CaWO$_4$ \cite{Antipin_1968} and the group around Patrice Bertet reported in more recent studies two EPR peaks with some similarities to the axial Er$^{3+}$ signal and a rhombic symmetry~\cite{Billaud.2025, LeDantec.2021}.
\section{Symmetry of reported EPR spectra}\label{S_Review}
In this study, we refer to these environments as sites. In literature, we found several historic examples of additional sites found in EPR measurements for different REIs incorporated into CaWO$_4$ \cite{Kedzie.1963, Nassau.1963, Ranon.1964, Mims.1967, Nemarich.1968, Garrett.1964}. Most of these studies reported a set of four lines, which transform into one another under a $\SI{90}{\degree}$ rotation by the c-axis \cite{Garrett.1964, Ranon.1964, Mims.1967} or did describe a similar behavior \cite{Kedzie.1963, Nassau.1963}. Mims and Gillen \cite{Mims.1967} described that there are two lines for such a compensation site in the a-b plane, whereas out of this plane the two lines split into four. Additionally, Kedzie and Kestigian \cite{Kedzie.1963} reported for the incorporation of Fe$^{3+}$ in CaWO$_4$ a single line for B$\parallel$c, which split up into four lines and merge into two lines for $\overset{\rightharpoonup}{\text{B}}\parallel [110]$. In the a-b plane ((001)-plane), the two lines merge into one line every $\SI{90}{\degree}$. They described the behavior of additional sites from Nd$^{3+}$ in a separate CaWO$_4$ crystal as similar to Fe$^{3+}$ resonances. Given the $\SI{90}{\degree}$ relation of reported compensation sites \cite{Garrett.1964, Ranon.1964, Mims.1967} the signal of the four inequivalent sites should also merge into one line for $\overset{\rightharpoonup}{\text{B}}\parallel$ c. Furthermore, the principal axes of these compensation sites are reported in \cite{Garrett.1964, Ranon.1964, Mims.1967}. We summarized the details in Table~\ref{T_A1_LitRev} as conclusive support for the following highlighting. One study describes the spectra as rhombic \cite{Ranon.1964} with orthogonal principal axes and without alignment with any crystal axis. Two of the four inequivalent sites own a principal axis along $[110]$ and the other along $[1\bar10]$. Nemarich and Viehmann \cite{Nemarich.1968} claimed that this rhombic spectrum is consistent with their spectra. It is important to note that Ranon and Voltera \cite{Ranon.1964} do not claim to have a rhombic crystallographic site; they only describe the spectra to be rhombic. Thus, we interpret that the local symmetry of their Nd$^{3+}$ plus compensation vacancy owns a rhombic symmetry tilted to the overall crystal lattice. Garrett and Merritt \cite{Garrett.1964} reported for vacancy-compensated Nd$^{3+}$ in CaWO$_4$ one set of four lines which own a not fully orthogonal principal axes with two principal axis projections into the a-b plane somewhat oriented towards the $[110]$ and $[1\bar10]$ axis. Mims and Gillen \cite{Mims.1967} conducted in this context a detailed study on Ce$^{3+}$ in CaWO$_4$ with and without Na$^{+}$ compensation. They reported several sets of additional sites for both cases. For the strongest vacancy ($I(\phi)$) and Na$^{+}$ ($I(Na)$) compensated spectra, they observed principal axes that are not fully orthogonal, and oriented somewhat towards the $[110]$ and $[1\bar10]$ direction. Furthermore, they found good agreement with predictions of the linear electric field for a nearest-neighbor compensation. For the second strongest observed vacancy spectra ($II(\phi)$), they reported orthogonal principal axes with some alignment to the a and b axes. However, they could not predict similar principal axes for this set and suggested that strain fields might need to be considered additionally. Tiranov et al. \cite{Tiranov.422025} reported three additional lower symmetry orthorhombic D2 sites in EPR measurements on $^{171}$Yb$^{3+}$:CaWO$_4$ and pointed towards nearby charge compensation centers.\\
Moreover, there are a different number of sets reported. For vacancy-compensated Nd$^{3+}$, there is only one set reported by \cite{Kedzie.1963} and \cite{Garrett.1964}. For vacancy-compensated Yb$^{3+}$, there is more than one additional set reported from \cite{Ranon.1964} and \cite{Nemarich.1968} and three sets by Tiranov et al. \cite{Tiranov.422025}. And \cite{Mims.1967} reported three sets for Ce$^{3+}$ with vacancy charge compensation and two sets with Na charge compensation, whereas \cite{Nassau.1963} reported two sets for vacancy-compensated Ce$^{3+}$. It is a bit surprising that favoring sites seems to somewhat correlate with the REI itself. However, this is not enough data to get a final conclusion. \\

Besides the charge imbalanced incorporation of REIs into CaWO$_4$, the Fe$^{3+}$:CaWO$_4$ system attracted some attention in literature too. For this system older studies reported orthorhombic spectra with similar symmetries as the REI incorporation and related it to incorporation at the W$^{6+}$ site \cite{Kedzie.1963, Kedzie.1965} or interstitial sites \cite{Golding.1978}. More recent studies, however, showed that Fe$^{3+}$ incorporates, like REIs, into the Ca$^{2+}$ site. In this site it does not exhibit an EPR signal unless charge compensation is introduced. In this case, the Fe$^{3+}$ showed four non-equivalent charge compensation sites with an orthorhombic spectrum similar to REI signals~\cite{LeDantec.2021,Billaud.2025,Claridge.1997}. Its comparably isotropic absorption corresponding to a g-factor of 4.3, can even be found in its naturally abundant form in highly purified CaWO$_4$ due to its sharp transition~\cite{LeDantec.2021,Billaud.2025}.\\

Optical investigations of the g-tensor of Er$^{3+}$ incorporated into different crystalographic sites in different systems were only reported in recent years \cite{Sun.2008, Bottger.2024, Bottger.2025, Holzapfel.2024, Vinh.2003}. These measurements bear additional obstacles compared to EPR measurements, as optical selection rules become relevant, and instead of a single transition, four transitions appear for two split degenerate states. Depending on the broken symmetry operations through the new crystallographic site, even additional transitions can appear.\\ \\

In this paper, we measure the Zeemann-split polarization dependent absorption and symmetry of our previously identified sites 1 - 3. Using measurements at different temperatures, we identify different potential sets of Zeeman-split transitions and highlight the similarities to EPR studies. Finally, we calculate the effective g-factors, compare them to literature, or to an EPR measurement.
\section{Sites}\label{S_Sites}
\begin{figure*}[t]
    \centering
    \includegraphics[page=1, width=\linewidth]{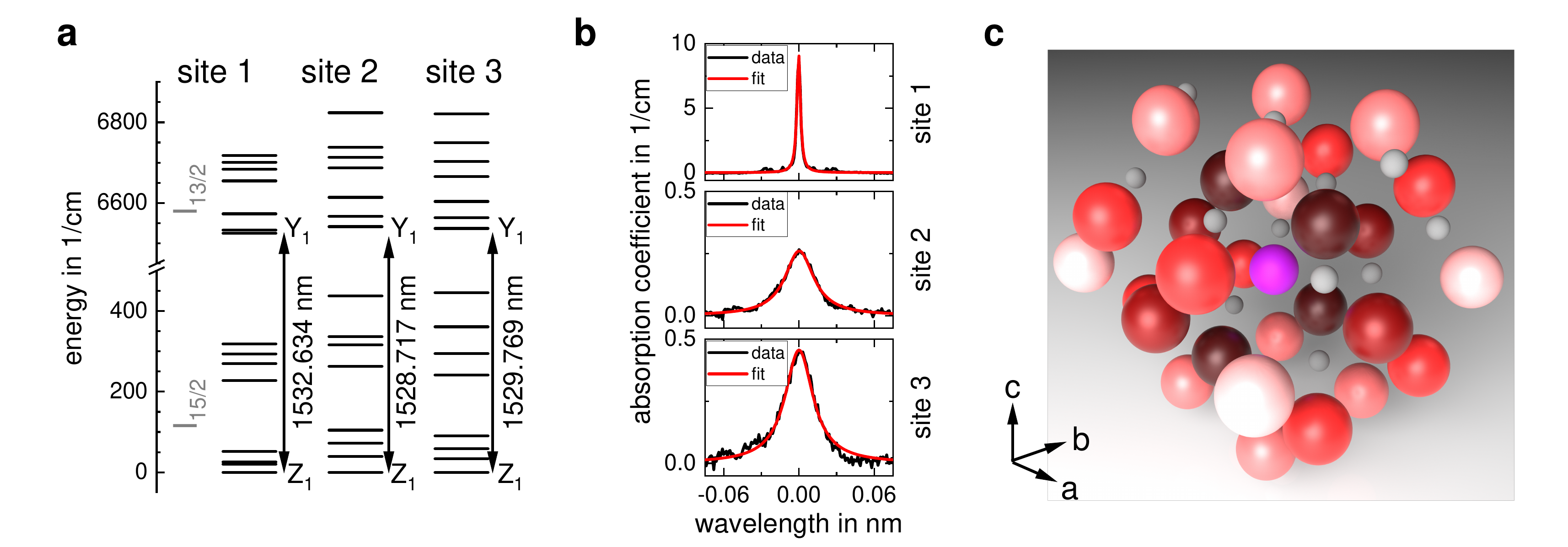}
    \caption{\textbf{a:} Energy levels of site 1 - 3 with highlighted investigated Z1Y1 transition. (energies taken from \cite{Becker.2025}) \textbf{b:} Absorption coefficients of sites 1 - 3 in $\alpha$-polarization with Lorentzian fit. \textbf{c:} CaWO$_4$ lattice presented in ionic radii surrounding an Er$^{3+}$ (violet) incorporated at Ca$^{2+}$ (reddish) site. Oxygen ions are excluded for visibility, and W$^{6+}$ ions are presented in grey. Ca$^{2+}$ is color-coded from dark red (nearest neighbor set) to bright red (5th nearest neighbor set)}
    \label{F_Figure_01}
\end{figure*}
In order to cover the background, we introduce the basic system in Figure \ref{F_Figure_01}. Figure \ref{F_Figure_01}a illustrates the $\mathrm{I}_{13/2}$ and $\mathrm{I}_{15/2}$ energy levels we identified previously \cite{Becker.2025}. In this study, we  only investigate the Zeeman symmetries of the highlighted Z1Y1 transitions. Figure \ref{F_Figure_01}b shows the absorption coefficient of these lines at zero magnetic field under $\alpha$-polarization (k$\parallel$c). Site 2 and 3 show a maximum of approximately $\SI{3}{\%}$ and $\SI{5}{\%}$ of site 1's absorption. Additionally, the inhomogeneous linewidths are $\Delta_{1,\text{inh.}}=\SI{430(7)}{MHz}$, $\Delta_{2,\text{inh.}}=\SI{3.57(8)}{GHz}$ and $\Delta_{3,\text{inh.}}=\SI{3.46(9)}{GHz}$. This is $\SI{18}{\%}$ to $\SI{22}{\%}$ smaller compared to our previous reported linewidths. We attribute this difference to power broadening as we recorded the absorption measurements in this study with laser power in the 10s of pW range, whereas the previous study was recorded at 100s of µW at the crystal. \\
As introduced in Section~\ref{S_Review}, additional sites were seen in EPR studies for other rare-earth ions doped into CaWO$_4$. These studies suggest a relation to nearby vacancies at neighboring Ca$^{2+}$ sites \cite{Nassau.1963, Ranon.1964, Mims.1967, Garrett.1964}. Figure \ref{F_Figure_01}c shows a reduced CaWO$_4$ environment of the Er$^{3+}$ incorporated at a Ca$^{2+}$ site in the ionic radius picture (data taken as in \cite{Becker.2025} from \cite{MatProj_CaWO4} and \cite{Haynes.op.2014}). The Ca$^{2+}$ ions of the five neighboring sets are color-coded from dark red to bright red, corresponding to the total distance to the Er$^{3+}$ ion. Geometrical details of the five depicted sets of Ca$^{2+}$ ions are summarized in Table~\ref{T_Geometry}.
\begin{table*}[ht]
\centering    
\begin{tabular}{
    >{\centering\arraybackslash}p{2.8cm} 
    !{\vrule width 1pt}!{\vrule width 1pt}
    >{\centering\arraybackslash}p{2.7cm} | >{\centering\arraybackslash}p{2.7cm} | >{\centering\arraybackslash}p{2.7cm} | >{\centering\arraybackslash}p{2.7cm} | >{\centering\arraybackslash}p{2.7cm}
}
\toprule
\makecell{ } & \makecell{Set 1} & \makecell{Set 2} & \makecell{Set 3} & \makecell{Set 4} & \makecell{Set 5} \\ \addlinespace[0.2em] 
\hline
\hline \addlinespace[0.2em] 
Color/Hue & \LARGE\textcolor[rgb]{0.31, 0.07, 0.09}{$\bullet\:$}\normalsize very dark red & \LARGE\textcolor[rgb]{0.6, 0.08, 0.08}{$\bullet\:$}\normalsize dark red & \LARGE\textcolor[rgb]{1, 0.22, 0.22}{$\bullet\:$}\normalsize medium red & \LARGE\textcolor[rgb]{0.96,0.45,0.45}{$\bullet\:$}\normalsize  bright red & \LARGE\textcolor[rgb]{1, 0.82, 0.82}{$\bullet\:$}\normalsize very bright red\\ [0.5em]
Number of Ions & 4 & 4 & 8 & 8 & 4 \\ [0.5em]
Distance in \AA & 3.87 & 5.26 & 6.53 & 6.78 & 7.43 \\ [0.5em]
\makecell{Angle to b-axis\\in a-b plane in $^\circ$} & 0/90 & 0/90 & 26.6/63.4 & 45 & 45 \\ [1.0em]
\makecell{Angle to c-axis\\in b-c plane in $^\circ$} & 0/42.8 & 90 & 61.6/42.8 & 24.9 & 90 \\
\bottomrule
\end{tabular}
\caption{Summary of geometrical orientations of neighboring Ca$^{2+}$ sites with respect to the Er$^{3+}$ incorporation site.}
\label{T_Geometry}
\end{table*}
In this simplified geometrical picture we draw a line between the neighboring Ca$^{2+}$ sites and the Er$^{3+}$ incorporation site. We project the angles of this line into the a-b and b-c planes and calculate the corresponding angles. Comparing these angles yields that set 4 and 5 are equally oriented in the a-b plane and own a different angle to the c-axis in the b-c plane. Set 1 owns four distinct angles, whereas set 2 is symmetric in the b-c plane and owns two angles in the a-b plane. Set 3 owns, similar to set 1, four distinct angles, however, none of these angles is aligned with any crystal axis. \\
For simplification, we only consider the five neighboring Ca$^{2+}$ site sets around the Er$^{3+}$ incorporation, as closer vacancies should provide a stronger perturbation. \\ \\
\begin{figure}[t]
    \centering
    \includegraphics[page=1, width=\linewidth]{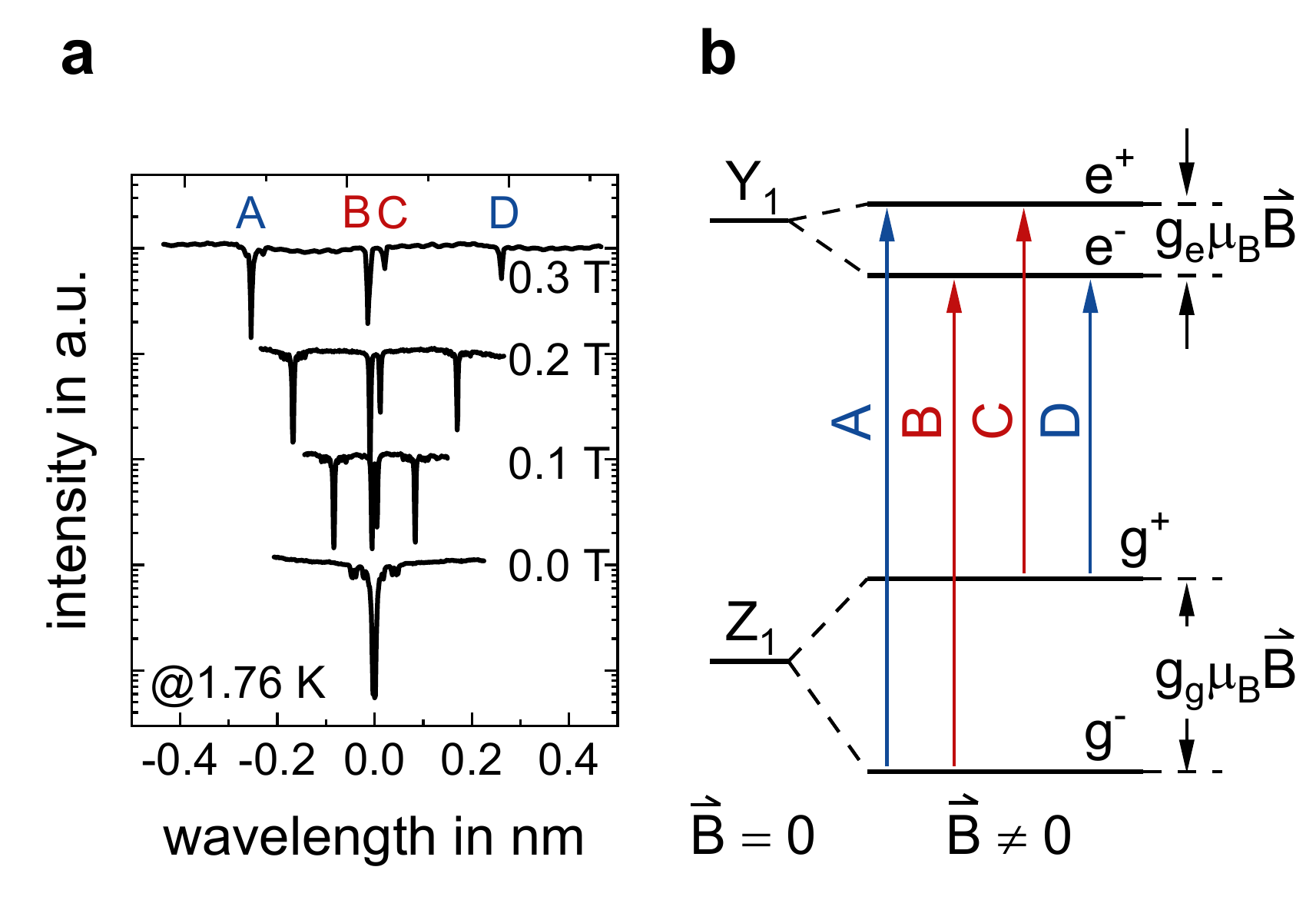}
    \caption{Zeeman splitting of a Z1Y1 transition \textbf{a:} Zeeman splitted Z1Y1 transition of site 1 at $\SI{1.76}{K}$ in $\alpha$-polarization with B$\parallel$b for different magnetic field magnitudes. \textbf{b:} Schematic Zeeman splitting of Z1Y1 transitions using the convention from \cite{Bottger.2009}.}
    \label{F_Figure_00}
\end{figure}
 Our here studied Z1Y1 system splits up into at least four optical transitions. Figure \ref{F_Figure_00}a shows an example absorption spectra of the Z1Y1 transition of site 1 with increasing magnetic field along the b-axis. With increasing magnetic field, the highlighted transitions A - D split. Additionally, the intensity of transitions C and D decreases. Figure \ref{F_Figure_00}b illustrates the involved energy levels and transitions using the convention from \cite{Bottger.2009}. With an increasing magnetic field, the previously degenerate levels of Y1 $\mathrm{e}^+$ and $\mathrm{e}^-$, as well of Z1 $\mathrm{g}^+$ and $\mathrm{g}^-$ split further up depending on the effective g-factor of the excited state $\mathrm{g}_e$ and ground state $\mathrm{g}_g$. This leads to an increasing spectral separation of transitions A - D. The spin-flipping transitions A and D can be identified as the transitions with the highest and lowest energy. However, depending on whether $\mathrm{g}_e$ or $\mathrm{g}_g$ is larger, the spin-preserving transitions B and C can not be identified by their energy relation, but by their population. The decreasing intensity of transition C and D in Figure \ref{F_Figure_00}b has its origin in a less and less populated $\mathrm{g}^+$ state by the Boltzmann distribution. This can be achieved by an increased level splitting or by a reduced temperature. \\
In case our sites own a reduced point group symmetry from S$_4$ of the Ca$^{2+}$ site, we would expect more transitions to appear \cite{Holzapfel.2024, Bottger.2006, Bottger.2024, Bottger.2025}. For the potential C$_2$ (C$_1$) point symmetry, we would expect two (four) magnetic inequivalent classes. Thus, we would expect 2x4 (4x4) optical absorbing transitions. In our recorded measurements, we could not identify more than four absorbing transitions for any site. However, this could come from a too low signal-to-noise ratio and/or forbidden transition selection rules rather than from an unreduced crystallographic point group.
\section{Measured Zeemann Symmetries} \label{Sec_Host}
\begin{table*}[t]
\renewcommand{\arraystretch}{1.4} 
\centering
\begin{tabular}{c|c|c|c|c|c|c|c}
\multirow{2}{*}{} & \multicolumn{5}{c|}{$g_g$} & \multicolumn{2}{c}{$g_e$} \\
\cline{2-8}
& fit & EPR \cite{Mason.1968}&EPR \cite{Antipin_1968}&Calc. \cite{Enrique.1971} & optical \cite{Ourari.192023} & fit & optical \cite{Ourari.192023} \\
\hline
B$\parallel$c ($g_\parallel$) & $\SI{1.263(10)}{}$ & $\SI{1.247(1)}{}$& $\SI{1.247(3)}{}$&$\SI{1.21}{}$& 1.4 & $\SI{1.453(9)}{}$ & 1.3 \\
B$\parallel$b ($g_\perp$)     & $\SI{8.42(3)}{}$& $\SI{8.400(3)}{}$&$\SI{8.38(2)}{}$&$\SI{8.45}{}$  & 8.6 & $\SI{7.44(2)}{}$ & 7.6 \\
\end{tabular}
\caption{Summary of fitted g factors of site 1 with literature values. $g_\parallel$ is fitted from an increasing magnetic field sweep instead of a rotation scan, to include paramagnetic shifts observed for $\overset{\rightharpoonup}{\text{B}}\parallel$ c.}
\label{T_S1_gFactor}
\end{table*}
\begin{figure}[t]
    \centering
    \includegraphics[page=1, width=\linewidth]{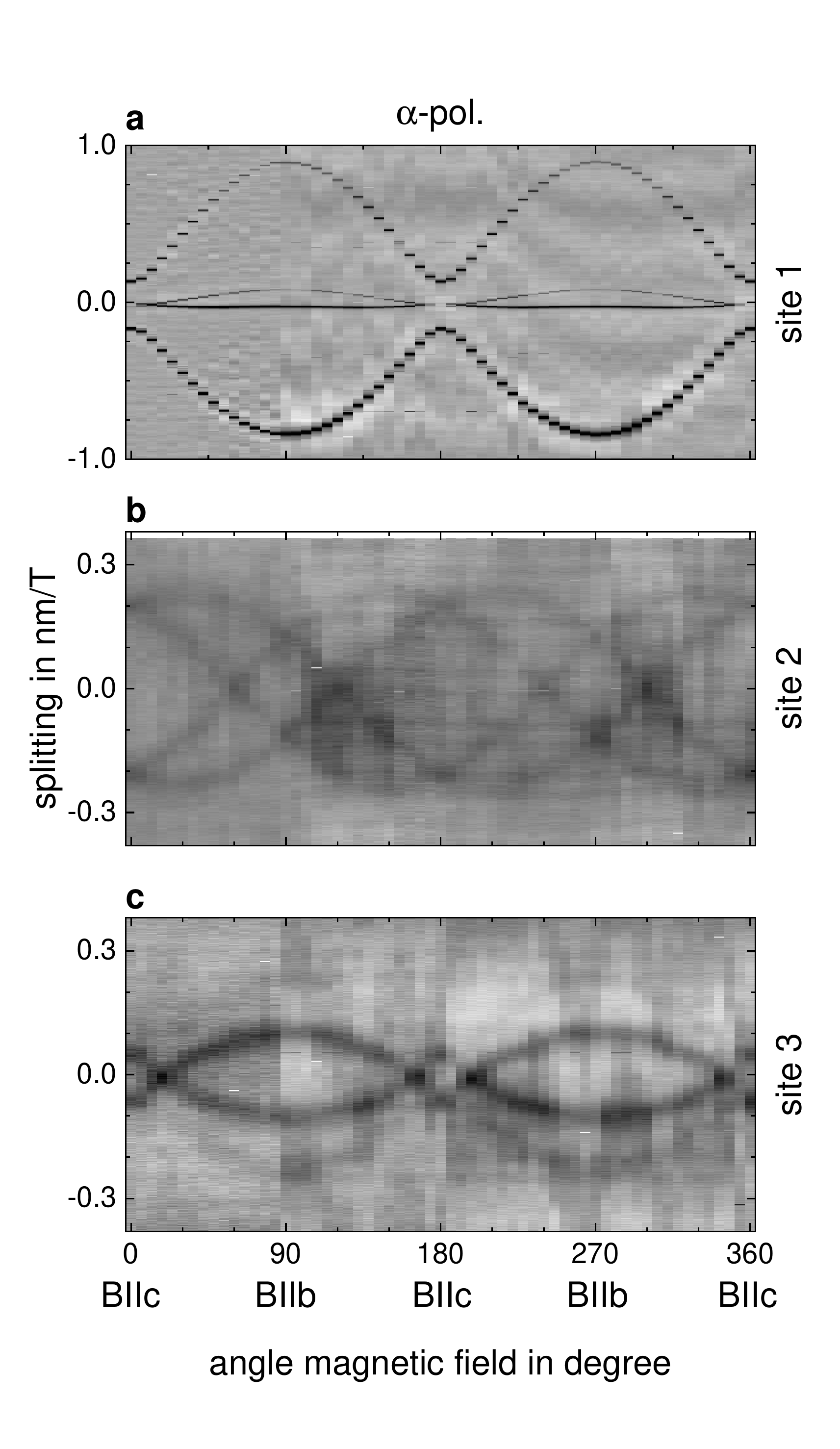}
    \caption{Symmetry of Zeeman splitted transitions of site 1 - 3 in the b-c plane measured in $\alpha$-polarization. Site 1 is measured at $\SI{0.6}{T}$ at $\SI{1.76}{K}$, site 2 and site 3 are measured at $\SI{0.5}{T}$ at $\SI{10}{K}$.}
    \label{F_Figure_02}
\end{figure}
\begin{figure}[t]
    \centering
    \includegraphics[page=1, width=\linewidth]{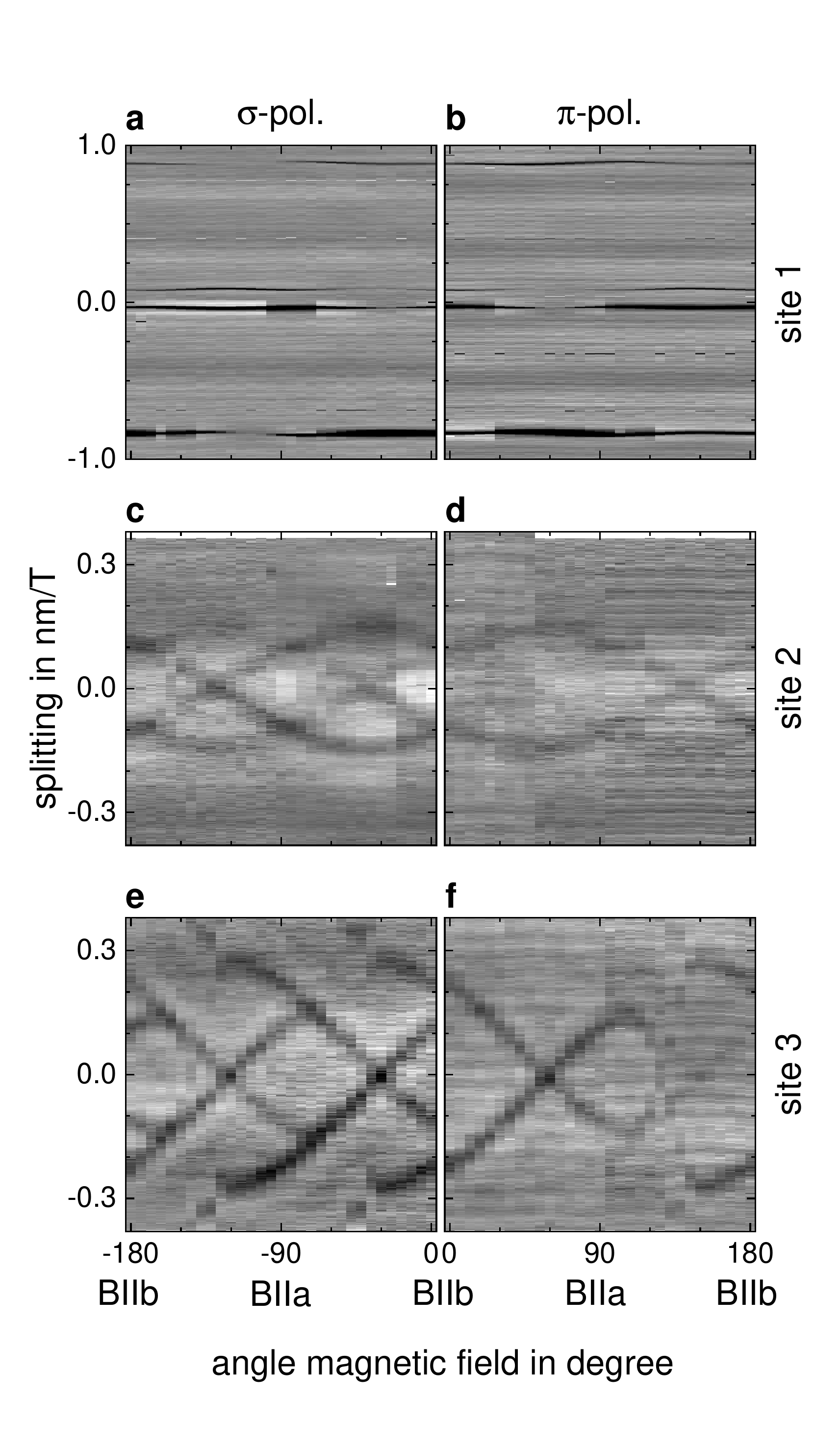}
    \caption{Symmetry of Zeeman split transitions of site 1 - 3 in the a-b plane measured in $\sigma$ ($\SI{-180}{\degree}$ to $ \SI{0}{\degree}$, \textbf{a, c, e}) and $\pi$ ($\SI{0}{\degree}$ to $\SI{180}{\degree}$, \textbf{b, d, f}) - polarization. Site 1 is measured at $\SI{0.6}{T}$ at $\SI{1.76}{K}$, site 2 and site 3 are measured at $\SI{0.5}{T}$ at $\SI{10}{K}$.}
    \label{F_Figure_03}
\end{figure}
Figure \ref{F_Figure_02} and \ref{F_Figure_03} show Zeeman split absorption transitions of sites 1 - 3 with the magnetic field rotated in the b-c and a-b plane under $\alpha$ (k$\parallel$c, E$\perp$c), $\sigma$ (k$\perp$c, E$\perp$c), and $\pi$-polarization (k$\perp$c, E$\parallel$c). We measured site 1 at a total magnetic field of $\SI{0.6}{T}$ at $\SI{1.76}{K}$, while we applied $\SI{0.5}{T}$ for site 2 and 3 at $\SI{10}{K}$. We choose these measurement conditions so that no Zeeman transitions of other sites interfere in these scans, the $g^{+}$ states are decently populated, and the magnetic field magnitude is large enough to avoid the mixed regime at low magnetic fields. For larger wavelength scans of sites 2 and 3, we could not identify any further transitions. The concluded and fitted transitions are presented in Figures \ref{F_Figure_02_Overlay} and \ref{F_Figure_03_Overlay}. \\
Site 1 shows a total of four transitions with a $\SI{180}{\degree}$ symmetry in the b-c plane, axial extremes, and a crossing at approximately $\pm\SI{12}{\degree}$ around the c-axis. For $\overset{\rightharpoonup}{\text{B}}\parallel$ c, the inner two transitions appear suppressed. In the a-b plane, site 1 has a minimal off-axis bending. Additionally, both polarizations show locally suppressed absorption lines. Furthermore, we can assign the transitions in the a-b plane bottom-up to the A, B, C, and D, and along the c-axis, to A, C, B, and D. To validate if our observed symmetries align with the literature, we calculated the g-factors for all magnetic field orientations with Equations \ref{E_gg} and \ref{E_ge}, and subsequently fitted them using Equation~\ref{E_geff}. Figure~\ref{F_Figure_A_02}a shows the detailed fit, and Table~\ref{T_S1_gFactor} summarizes the resulting g factors.
\begin{align}
    g_g &=\frac{1}{\mu_B \overset{\rightharpoonup}{\text{B}}}\cdot\frac{1}{2}\cdot\left(\left(E_A-E_C\right)+\left(E_B-E_D\right)\right)\label{E_gg}\\
    g_e &=\frac{1}{\mu_B \overset{\rightharpoonup}{\text{B}}}\cdot\frac{1}{2}\cdot\left(\left(E_A-E_B\right)+\left(E_C-E_D\right)\right)\label{E_ge}\\
    g_{\text{eff}}&=\sqrt{\left(g_\perp \sin\theta\right)^2+\left(g_\parallel \cos\theta\right)^2} \label{E_geff}
\end{align}
Our fitted values for $g_{g}$ align well with literature values from EPR measurements \cite{Mason.1968,Antipin_1968} and calculation \cite{Enrique.1971}. However, there is a deviation to reported values measured optically \cite{Ourari.192023}, likely as the authors mentioned, due to the uncertainty of their vector magnet calibration. Additionally, we identified for B$\parallel$c that transition C is at larger energies than B, leading to $g_{g,\parallel}<g_{e,\parallel}$.\\ \\
Site 2 shows a total of four transitions, which merge into two transitions along the crystal axes. Additionally, an anti-crossing in the b-c plane at approximately $\pm \SI{30}{\degree}$ around the b-axis and an anti-crossing in the a-b plane between $\SI{-36}{\degree}$ and $\SI{-42}{\degree}$ as well as every additional $\SI{90}{\degree}$ is present. We identified the anti-crossing by comparing these scans at $\SI{10}{K}$ with scans at $\SI{1.76}{K}$ (see Figure \ref{F_Figure_A_01}a), which indicates that the lower transitions are connected to the $g^{-}$ state and the upper transitions to the $g^{+}$ state. Furthermore, the transitions extreme in the b-c plane are around the c axis between $\pm \SI{30}{\degree}$ and $ \SI{36}{\degree}$ and in the a-b plane above the anti-crossings. Additionally, under $\pi$-polarization, it seems that one set of transitions is favored. Given the anti-crossings and polarization dependence, we conclude that we see two sets of two transitions connecting the different ground states with the same excited state. For instance, it could be two sets of A and C or B and D transitions. Using this relation and the fitted transitions visible in Figure \ref{F_Figure_02_Overlay} and \ref{F_Figure_03_Overlay}, we calculated the effective g factors of these scans and present them in Figure \ref{F_Figure_A_02}. According to this, we get $g_g,\perp\approx\SI{2}{}$ and $g_g,\parallel\approx\SI{3.7}{}$, which aligns with the note of different g-factors of axial and compensation sites in \cite{Antipin_1968}. However, when we compare the relative difference of principal g-factors of other vacancy-compensation sites from Table \ref{T_A1_LitRev} there is some inconsistency. For instance, sites 2 $g_3$ would be close to 4, which is quite different from the axial 1.263, while other reported $g_3$ values are similar to their axial site. Additionally, we identified with an EPR scan for $\overset{\rightharpoonup}{\text{B}}\parallel$ b a peak at 8.4 corresponding to $g_g$ of site 1 and a dip approximately at a g-factor of 2, which could be $g_g$ of site 2. Finally, if we compare these symmetry considerations with the naive geometrical picture drawn in Section \ref{S_Sites}, only set 3 has non-axial relations to the Er$^{3+}$ incorporation site. Especially the different and non-axial orientations of anti-crossings and extremes in the b-c plane require a geometrical relation similar to set 3. \\ \\
Site 3 shows also a total of four transitions, which merge into two transitions for $\overset{\rightharpoonup}{\text{B}}\parallel$ c. The outer transitions in the b-c plane appear less pronounced compared to the inner two. In contrast, the $\sigma$-polarization in the a-b plane shows all transitions with a similar intensity. Additionally, a crossing is visible in the b-c plane at approximately $\pm\SI{18}{\degree}$ off the c-axis. In the a-b plane four main crossings are visible at $\SI{-120}{\degree}$ and every additional $\SI{90}{\degree}$. We identified the crossing again by comparing the measurement at a lower temperature (see Figure \ref{F_Figure_A_01}b). Furthermore, this yields that for $\overset{\rightharpoonup}{\text{B}}\parallel$ a or $\overset{\rightharpoonup}{\text{B}}\parallel$ b the lowest and the third lowest transitions belong to the $g^{-}$ state while the other two transitions belong to the $g^{+}$ state. Moreover, in the $\pi$-polarization the upper and lower transitions cross sharply at around $\SI{120}{\degree}$, while this crossing is not apparent in the $\sigma$-polarization. Furthermore, only half of the transitions are clearly visible in the $\pi$-polarization.\\
From symmetry and polarization we conclude that again, two times the same two transitions are present with a different oriented dipole for the $\pi$-polarization. Both sets of two transitions merge to one another under a $\SI{90}{\degree}$ rotation in the a-b plane. However, in the b-c plane, they seem to reach a maximum of distinction for $\overset{\rightharpoonup}{\text{B}}\parallel$ b, while they become indistinguishable for $\overset{\rightharpoonup}{\text{B}}\parallel$ c. This indicates that the two sets are perpendicular to each other, with one oriented more along the a-axis and the other more along the b-axis. The crossings also suggest that the two involved transitions have, in addition to a different ground state, a different excited state. For instance, the transition combinations A and D or B and C are possible. However, the combination of A and D would require that all transitions merge into one, which we can not observe. Hence, the visible transitions must be two sets of the B and C transitions. With this $g_e > g_g$ for $\overset{\rightharpoonup}{\text{B}}\parallel$ b and $g_g > g_e$ for $\overset{\rightharpoonup}{\text{B}}\parallel$ c explain the curves and clean crossing in the b-c plane. Finally, from the naive geometrical picture, sets 1 and 2 have each two sites along and perpendicular to the b-axis in the a-b plane. However, in the b-c plane, set 2 has one angle to the c-axis while set 1 has two different angles. Unfortunately, both orientations can, in principle, cause a symmetry as seen in the b-c plane.

\section{Discussion}
In general, both site 2 and 3 show absorption spectra with symmetries similar to reported REIs or Fe$^{3+}$ incorporation in CaWO$_4$ experiencing charge compensation at nearby Ca$^{2+}$ sites (see Section \ref{S_Review}).\\
Both sets of site 2 repeat in a $\SI{180}{\degree}$ symmetric fashion in b-c and a-b plane. Thus, we would conclude a rhombic spectrum. As we are not certain about the orthogonality of their principal axis, a rhombic-like spectrum is likely the most precise description. However, given that the spectra are not aligned with any crystal axis, the crystallographic point group of the Er$^{3+}$-vacancy complex is probably C1.\\
The two sets of site 3 show a $\SI{180}{\degree}$ symmetry in the b-c plane and are aligned with the crystal axis. This indicates that one principal axis is aligned along [001], whereas the other two lie in the (001) plane. As site 3, however, shows a sawtooth like asymmetry, it does not agree as well as site 2 with a rhombic spectrum. Nevertheless, there is still some rhombic character in the spectrum. For the crystallographic point group, we conclude as well C1.\\
For comparison, we added the predictions of our principal axis to the literature summary in Table \ref{T_A1_LitRev}.\\
Additionally, a slight misalignment of the crystal with respect to the magnetic coils of up to $\SI{5}{\degree}$ is likely for the a-b plane scans and could explain the wiggles of our axial site present in Figure \ref{F_Figure_03}. \\
Finally, we could not observe non-zero field splitting of sites 2 or 3. In contrast, the two not yet identified but Er$^{3+}$ like peaks (b1 and b2) from \cite{LeDantec.2021} and \cite{Billaud.2025} seems to own a non-zero field splitting. Thus, we believe that the origin of our sites and the EPR peaks reported in \cite{Billaud.2025, LeDantec.2021} is different.\\ \\
To conclude, we found good agreement of rhombic-like spectra at our sites 2 and 3 with literature findings. As the large majority of these findings suggest certain neighboring Ca$^{2+}$ vacancy sites, we conclude that our sites are also Er$^{3+}$-vacancy complexes. Additionally, in the naive geometrical picture, we suggest that our site 2 corresponds to the third nearest neighbor and site 3 to the nearest neighbor or next-nearest neighbor Ca$^{2+}$ site vacancy set. Finally, we could work out some projections of principal axes and g-factors.

\section{Acknowledgements}\label{S_Acknowledge}
We gratefully acknowledge support
from the German Federal Ministry of Research, Technology and Space (BMFTR) via the project 6G-life, the Bavarian State Ministry for Science and Arts (StMWK) via project NEQUS, the Bavarian Ministry of Economic Affairs (StMWi) via project 6GQT, as well as from the German Research Foundation (DFG) under Germany’s Excellence Strategy EXC-2111 (390814868) and projects PQET (INST 95/1654-1) and MQCL (INST 95/1720-1).

\section{Data Availability}
The data that support the findings of this study are available from the corresponding author upon reasonable request.

\section{Appendix}
\subsection{Sample and Setup}
The measurements are performed on a $\SI{10}{ppm}$ Er$^{3+}$ doped CaWO$_4$. The CaWO$_4$ crystal was grown by a hybrid Czochralski-Flow-Zone approach by SurfaceNet. The measurements were taken inside an AttoDry2100 cryostat with a 3-9 T vector magnet. The sample was mounted on top of an in-house fabricated mirror with a reflectivity of approximately $\SI{95}{\%}$ at room temperature. The laser was stabilized using a wavemeter (HighFinesse WS7IR). The collimated laser was guided through the sample, reflected back at the mirror, guided back through the sample, and coupled to a polarization maintaining fiber. The fiber coupled signal was split into four channels of an SNSPD system from Single Quantum. To further improve our noise level, we installed a noise and blackout box around the cryostat, which additionally shielded the setup from air fluxes and kept it at an equilibrium temperature. 
\begin{table*}[h]
\centering
\begin{tabular}{l|cccc|cc|ccc|c|ccc}
\multirow{2}{*}{} & \multicolumn{4}{c|}{Ce$^{3+}$\cite{Mims.1967}} & \multicolumn{2}{c|}{Nd$^{3+}$\cite{Garrett.1964}}& \multicolumn{3}{c|}{Yb$^{3+}$\cite{Ranon.1964}}&Fe$^{3+}$\cite{Claridge.1997}& \multicolumn{3}{c}{Er$^{3+}$ this study}\\
\cline{2-14}
 & Axial & I ($\varphi$) & II ($\varphi$) & I (Na)\textsuperscript{b} & Axial & I ($\varphi$)& Axial & I ($\varphi$)& II ($\varphi$)&I ($\varphi$)&Axial & site 2 ($\varphi$)& site 3 ($\varphi$)\\
\hline
$g_1$ & 1.43 & 0.82 & 0.83 & 1.15 &2.54&3.11&3.920&4.788&3.704&4.300&8.42&&\\
$\theta_1$ & 90$^\circ$ & 84$^\circ$ & 90$^\circ$ & 87$^\circ$ &90$^\circ$&52$^\circ$&90$^\circ$&90$^\circ$&90$^\circ$&3$^\circ$&90$^\circ$&&90$^\circ$\\
$\varphi_1$ &  & 40$^\circ$ & 6$^\circ$ & 46$^\circ$ &&39$^\circ$&&45$^\circ$&45$^\circ$&40$^\circ$&&45$^\circ$&\\
$g_2$ & 1.43 & 1.61 & 1.75 & 1.57 &2.54&2.36&3.920&3.012&4.155&4.289&8.42&&\\
$\theta_2$ & 90$^\circ$ & 74$^\circ$ & 60$^\circ$ & 84$^\circ$ &90$^\circ$&74$^\circ$&90$^\circ$&84$^\circ$&83$^\circ$&7$^\circ$&90$^\circ$&&90$^\circ$\\
$\varphi_2$ &  & 131$^\circ$ & 96$^\circ$ & 136$^\circ$ &&142$^\circ$&&135$^\circ$&135$^\circ$&130$^\circ$&&135$^\circ$&\\
$g_3$ & 2.92 & 2.86 & 3.08 & 2.92 &2.03&1.54&1.058&0.975&1.155&4.288&1.263&&\\
$\theta_3$ & 0$^\circ$ & 17$^\circ$ & 30$^\circ$ & 7$^\circ$ &0$^\circ$&43$^\circ$&0$^\circ$&6$^\circ$&7$^\circ$&275$^\circ$&0$^\circ$&30$^\circ$&0$^\circ$\\
$\varphi_3$ &  & 290$^\circ$ & 276$^\circ$ & 293$^\circ$ &&250$^\circ$&&315$^\circ$&315$^\circ$&92$^\circ$&&&\\
note &tetra & & rhom?&tetra&& &tetra&rhom&rhom&o.rhom&tetra&rhom&rhom?\\
\hline
\end{tabular}
\caption{Summary of $g$ parameters for axial and nonaxial sites for different REIs in CaWO$_4$. Sites I are stronger compared to sites II. Used convention from \cite{Mims.1967}: $\theta_i$ denotes the angle between a principal axis $g_i$ and the $c$ axis, while $\varphi_i$ specifies the angle between the $c$ axis and the projection of $g_i$ onto the $ab$ plane. Each site produces four lines; the parameters of the other three lines are obtained by adding integer multiples of 90$^\circ$ to the corresponding set of angles $i$.}
\label{T_A1_LitRev}
\end{table*}

\subsection{Temperature Dependence}
\begin{figure}[h]
    \centering
    \includegraphics[page=1, width=\linewidth]{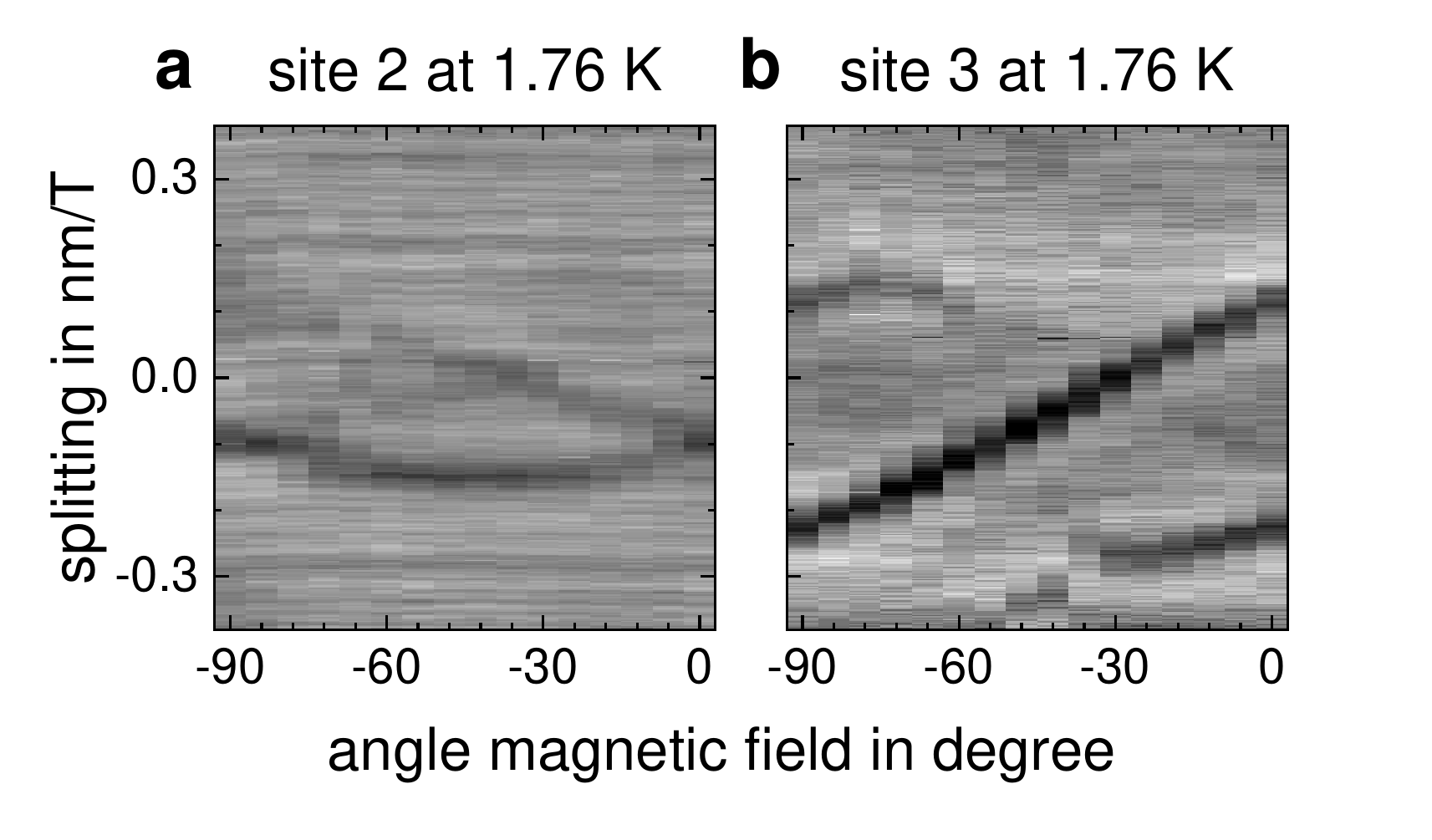}
    \caption{Symmetry of Zeeman split transitions of site 2 (\textbf{a}) and 3 (\textbf{b}) in the a-b plane measured in $\sigma$-polarization ($\SI{-90}{\degree}$ to $ \SI{0}{\degree}$) at $\SI{0.5}{T}$ at $\SI{1.76}{K}$.}
    \label{F_Figure_A_01}
\end{figure}

Figure \ref{F_Figure_A_01} shows additional rotation scans for site 2 and 3 in $\sigma$-polarization at $\SI{1.76}{K}$. In combination with the scans at $\SI{10}{K}$ in Figure \ref{F_Figure_03}c and e we can distinguish between transitions connected to different ground states. More on this in the main text.

\subsection{Color Map with fitted Transitions}
\begin{figure}[t]
    \centering
    \includegraphics[page=1, width=\linewidth]{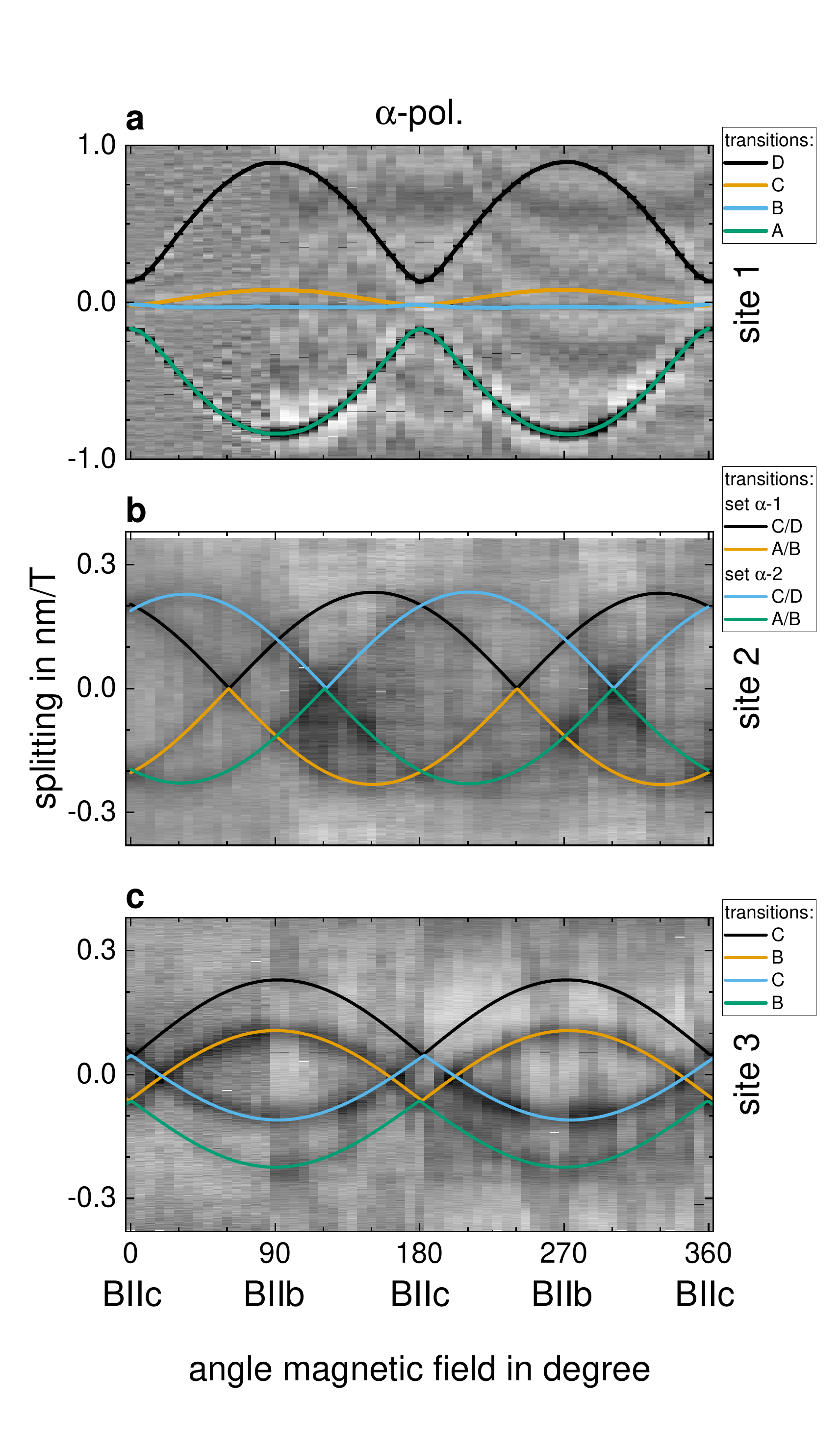}
    \caption{Overlayed Figure \ref{F_Figure_02} with fitted and assigned optical transitions.}
    \label{F_Figure_02_Overlay}
\end{figure}

\begin{figure}[t]
    \centering
    \includegraphics[page=1, width=\linewidth]{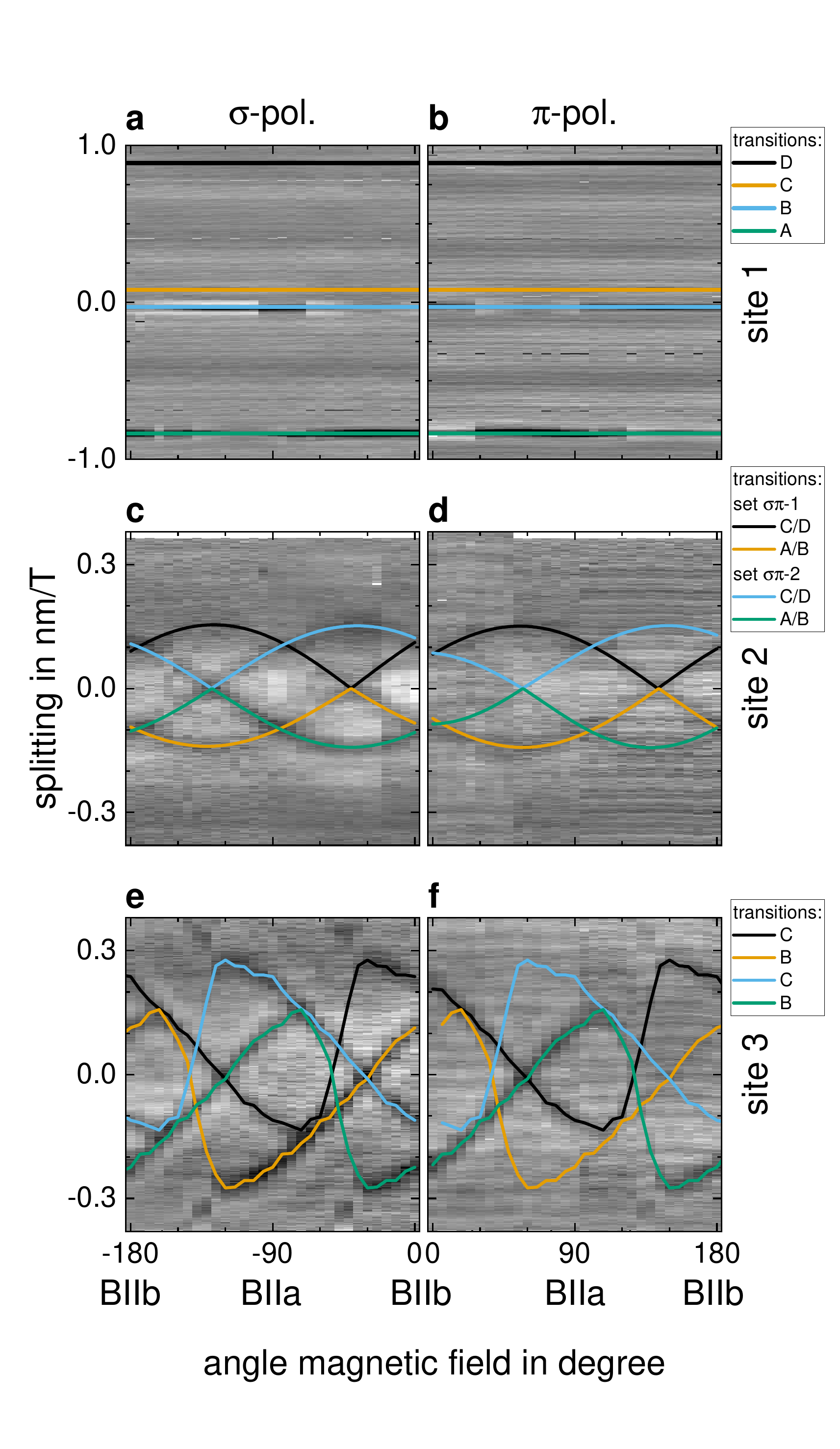}
    \caption{Overlayed Figure \ref{F_Figure_03} with fitted and assigned optical transitions.}
    \label{F_Figure_03_Overlay}
\end{figure}

In order to provide more clarity about our interpretations, we added fitted transitions or transition traces as overlays to Figure \ref{F_Figure_02} and \ref{F_Figure_03} and present them in Figure \ref{F_Figure_02_Overlay} and \ref{F_Figure_03_Overlay}. For site 1, we identified transitions A - D. For site 2, we identified two sets of either C and A or D and B. Additionally, we can not identify which set in the $\alpha$-polarization belongs to which set in the $\sigma$ or $\pi$-polarization, thus, we named them set $\alpha$-1, set $\alpha$-2, set $\sigma\pi$-1 and set $\sigma\pi$-2. For site 3, we also identified two sets of C and B transitions. However, we can not with certainty assign which C and B transitions belong together. We believe that transitions C and B belong together, which shows the extremes at the same orientation. This would mean black C and green B as well as cyan C and orange B belong together.
\subsection{Effective g-factor Fits}
\begin{figure}[t]
    \centering
    \includegraphics[page=1, width=\linewidth]{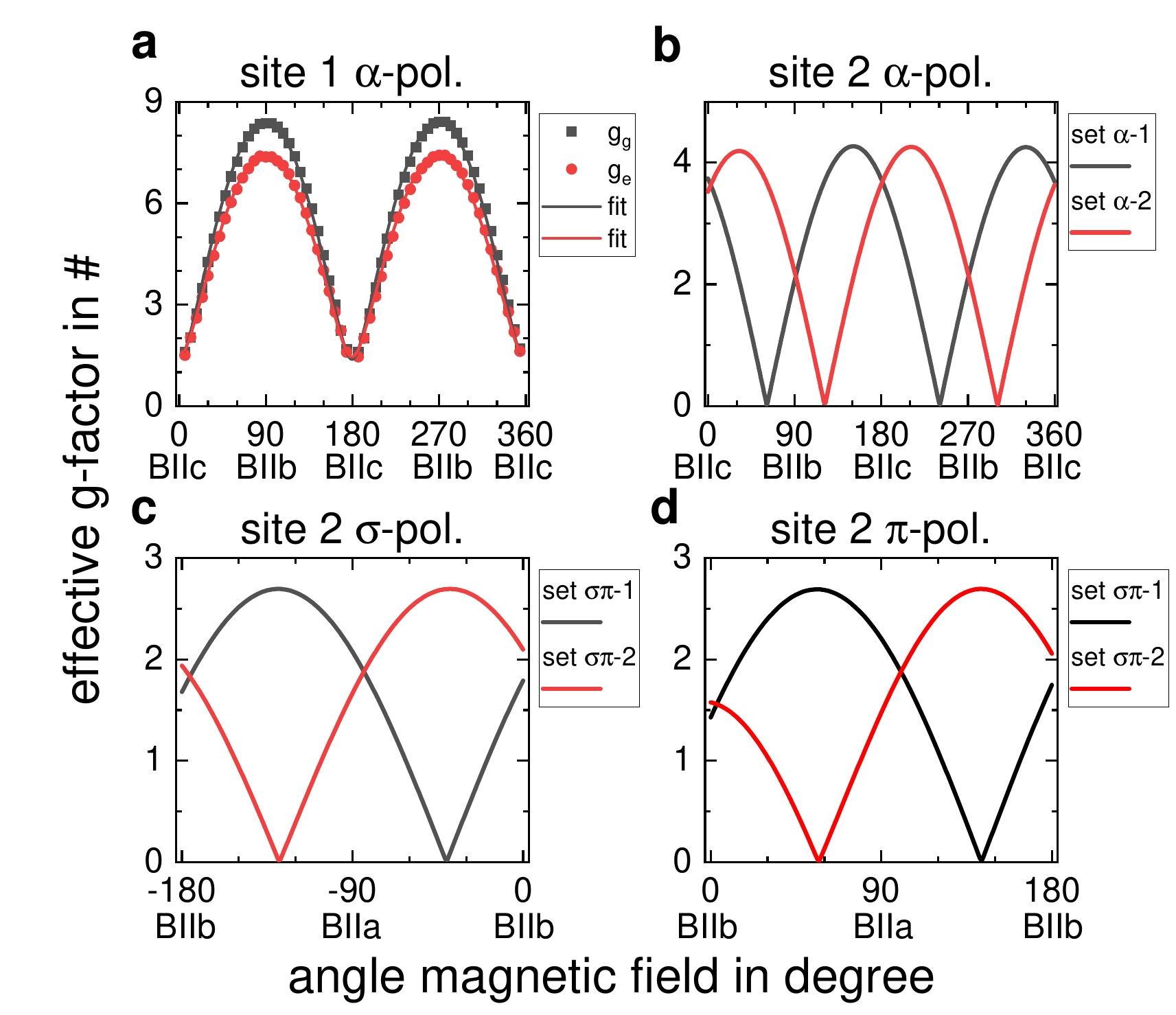}
    \caption{Fitted and calculated effective g-factors. \textbf{a} Site 1 with $g_g$ and  $g_e$. \textbf{b, c and d} Site 2 with $g_g$.}
    \label{F_Figure_A_02}
\end{figure}

\begin{figure}[t]
    \centering
    \includegraphics[page=1, width=\linewidth]{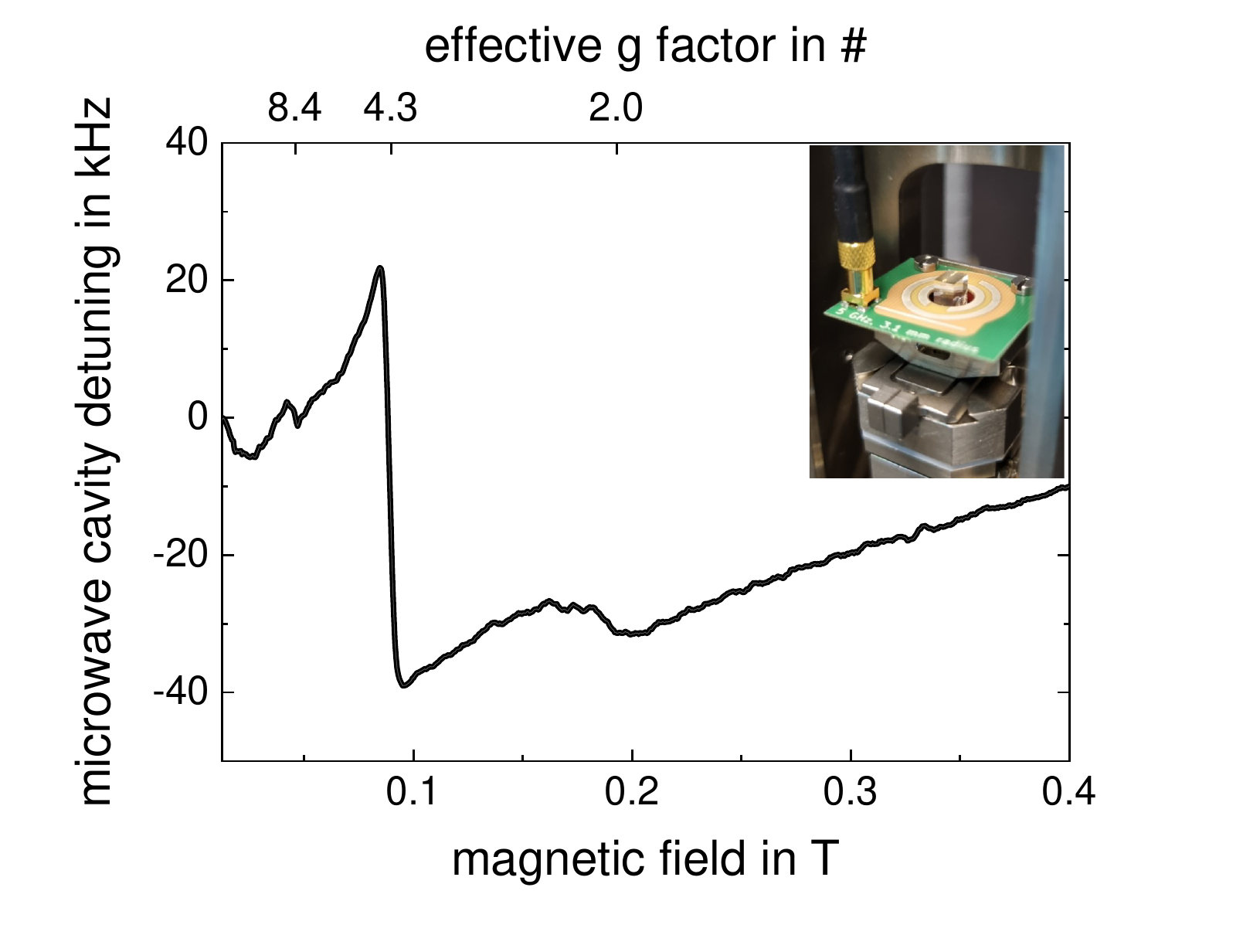}
    \caption{EPR measurement recorded at $\SI{1.76}{K}$ for $\overset{\rightharpoonup}{\text{B}}\parallel$ b at $f_0=\SI{5.402}{GHz}$ with a microwave cavity of $Q=\SI{270}{}$. The inset shows the crystal inside the PCB microwave cavity on top of our cryogenic piezoelectric stage tower.}
    \label{F_Figure_A_03}
\end{figure}

In Figure \ref{F_Figure_A_02} we provide the effective g-factors of site 1 ground and excited state as well as the effective ground state g-factor of site 2, as interpreted in the main text. For site 1 we used the transitions identified under $\alpha$-polarization and fitted with Equation \ref{E_geff}. For site 2 we used the fitted transition trace and calculated according to Equation \ref{E_gg} the corresponding $g_g$. Set $\sigma\pi$-2 of Figure \ref{F_Figure_A_02}d has only a small data set behind the transition fits; thus, its effective g-factors are less accurate compared to the other sets.
In addition to the optical measurements, we used a microwave cavity to investigate the ground state splitting using EPR measurements. Figure \ref{F_Figure_A_03} shows such a EPR measurement for B$\parallel$b with the inset showing the setup. In general, we can identify features at approximately an effective g-factor of 8.4, 4.3 and 2.0. The g-factor of 8.4 aligns with $g_{g,\perp}$ of site 1 (see Table \ref{T_S1_gFactor}). The g-factor of 2.0 is around what we expect for $g_{g,\perp}$ of site 2 according to Figure \ref{F_Figure_A_02}. The effective isotopic g-factor of 4.3 aligns well with the signature of  Fe$^{3+}$ in a Ca$^{2+}$ site with vacancies at neighboring Ca sites \cite{LeDantec.2021, Billaud.2025, Claridge.1997}, as outlined in Section~\ref{S_Review}. The purity of our CaWO$_4$ is with \SI{99.9999}{at\%} even better compared to the precursors CaCO$_3$ \SI{99.999}{at\%} and WO$_3$ \SI{99.998}{at\%}\cite{Erb.2013} of the study reported the outstanding electron spin coherence time of $\SI{23}{ms}$ \cite{LeDantec.2021}. On this undoped crystal, they reported a ratio of [Fe]/[Er] of 16 from EPR measurements. The lower purtiy of our doping precursor Er$_2$O$_3$ with \SI{99.9}{at\%} should be neglectibel, as only $\SI{10}{ppm}$ of the Er$^{3+}$ is doped. Thus, it seems likely that our Fe-peak originates as well from the few remaining Fe atoms with vacancy compensation centers close to.    

\bibliographystyle{apsrev4-1}

\providecommand{\noopsort}[1]{}\providecommand{\singleletter}[1]{#1}%

\end{document}